\documentclass[letterpaper, 10 pt, conference]{ieeeconf}  

\IEEEoverridecommandlockouts                              
\usepackage{graphicx} 
\usepackage{epstopdf} 
\usepackage{mathptmx} 
\usepackage{times} 
\usepackage{amsmath} 
\usepackage{amssymb}  
\usepackage{color}

\DeclareMathAlphabet{\mathcal}{OMS}{cmsy}{m}{n}
\usepackage{mathrsfs} 

\graphicspath{{img/}}

\newcommand{\M}{\mathcal{M}}

\newcommand{\R}{\mathbb{R}}
\newcommand{\U}{\mathcal{U}}
\newcommand{\Y}{\mathcal{Y}}

\newtheorem{proposition}{Proposition}
\newtheorem{remark}{Remark}

\newtheorem{theorem}{Theorem}
\newtheorem{definition}{Definition}
\newtheorem{corollary}{Corollary}
\newtheorem{example}{Example}

\graphicspath{{img/}}

\title{\LARGE \bf
Single module identifiability in linear dynamic networks
\thanks{This project has received funding from the European Research Council (ERC), Advanced Research Grant SYSDYNET, under the European Union's Horizon 2020 research and innovation programme (grant agreement No 694504).}%
}

\author{Harm Weerts, Paul M.J. Van den Hof and Arne Dankers
\thanks{Harm Weerts and Paul M.J. Van den Hof are with the Dept. of Electrical Engineering, Eindhoven University of Technology, The Netherlands {\tt \small h.h.m.weerts, p.m.j.vandenhof@tue.nl}}
\thanks{Arne Dankers is with the Electrical and Computer Engineering Dept. at the University of Calgary, Canada, {\tt\small adankers@hifieng.com}}%
}

\begin{document}

\maketitle
\thispagestyle{empty}
\pagestyle{empty}

\begin{abstract}
	A recent development in data-driven modelling addresses the problem of identifying dynamic models of interconnected systems, represented as linear dynamic networks.
	For these networks the notion network identifiability has been introduced recently,
	which reflects the property that different network models can be distinguished from each other.
	Network identifiability is extended to cover the uniqueness of a single module in the network model.
	Conditions for single module identifiability are derived and formulated in terms of path-based topological properties of the network models.
\end{abstract}

\section{Introduction}
Systems in engineering are becoming increasingly complex and interconnected. 
In many control, monitoring and optimization applications it is advantageous to model a system as a set of interconnected modules.
\emph{Linear dynamic networks} are formed by interconnecting modules according to a structured topology. 
Given the increasing availability of sensors, it is attractive to develop tools for data-driven modelling of linear dynamic networks.
There are several interesting topics of research, including the development of methods to estimate
the dynamics of one, several or all modules embedded in the network from a given data set, 
or estimating its topology \cite{Goncalves08,Materassi2010,Yuan2011,ChiusoPAuto2012}.

The objective in data driven modelling is to select a model that best represents the data from a set of candidate models, for example by using the setup introduced in \cite{VandenHof&Dankers&etal:13,Dankers_diss}.
When identifying either a full network or a subnetwork, it is important that the candidate models can be distinguished from each other.
For this purpose, the concept of network identifiability has been introduced in\cite{weerts2018identifiability}, as a follow up on system theoretic results of \cite{Goncalves08}. 
In this setting, network identifiability is dependent on the presence and location of external excitation signals, on structural information on the network topology and the disturbance correlation structure.

The analysis in \cite{weerts2018identifiability} has been concentrated on identifiability of a full network on the basis of all node variables being measured.
An alternative problem is formulated in \cite{bazanella2017identifiability} where identifiability of all or some of the modules is studied on the basis of a subset of node signals being measured. 
The results in \cite{bazanella2017identifiability} are formulated for the particular situation that all nodes are excited by external excitation signals and that the network is noise-free. 
Moreover, it is shown that the conditions for identifiability can be recast into attractive path-based conditions if the concept of network identifiability is considered in a generic sense.

In this paper we will extend the identifiability analysis of \cite{weerts2018identifiability} in two different ways. 
First we will cover identifiability of a single module in a noise disturbed network where all nodes are measured, in other words under what conditions is a module of interest identifiable?
And secondly we will show that, in line with the approach in \cite{bazanella2017identifiability}, the conditions for identifiability can be formulated as path-based conditions, if we consider the identifiability concept in a generic sense. 
This allows for a simple verification of identifiability based on the topology of the network models.

In Section \ref{sect:netw} the problem setup is provided and we recall the main results of \cite{weerts2018identifiability}. 
Next the concept of single module identifiability is defined, after which necessary and sufficient conditions are formulated for this type of network identifiability. 
Then it is shown that these conditions can be formulated as path-based conditions on the network, for the situation that we accept network identifiability as a generic concept.

\section{Preliminaries}
\subsection{Network setup}\label{sect:netw}
Following the basic setup of \cite{VandenHof&Dankers&etal:13}, a dynamic network is built up out of $L$ scalar \emph{internal variables} or \emph{nodes} $w_j$, $j
= 1, \ldots, L$, and $K$ \emph{external variables} $r_k$, $k=1,\ldots K$.
Each internal variable is described as:
\begin{align}
w_j(t) = \sum_{\stackrel{l=1}{l\neq j}}^L
G_{jl}(q)w_l(t) + \sum_{k=1}^K
R_{jk}(q)r_k(t) + v_j(t)
\label{eq:netw_def}
\end{align}
where $q^{-1}$ is the delay operator, i.e. $q^{-1}w_j(t) = w_j(t-1)$;
\begin{itemize}
	\item $G_{jl}$, $R_{jk}$ are proper rational transfer functions, and the single transfers $G_{jl}$ are referred to as {\it modules}.
	\item $r_k$ are \emph{external variables} that can directly be manipulated by the user;
	\item $v_j$ is \emph{process noise}, where the vector process $v=[v_1 \cdots v_L]^T$ is modelled as a stationary stochastic process with rational spectral density, such that there exists a $p$-dimensional white noise process $e:= [e_1 \cdots e_p]^T$, $p \leq L$, with covariance matrix $\Lambda>0$ such that
\[ v(t) = H(q)e(t). \]
\end{itemize}
For $p=L$, $H$ is square, stable, monic and minimum-phase.
The situation $p < L$ is referred to as the \emph{rank-reduced} noise case, and for a detailed description we refer to \cite{weerts2018identifiability}.
%
%

When combining the $L$ node signals we arrive at the full network expression
\begin{align*}
\begin{bmatrix}  \! w_1 \!  \\[1pt] \! w_2 \!  \\[1pt]  \! \vdots \! \\[1pt] \! w_L \!  \end{bmatrix} \!\!\! = \!\!\!
\begin{bmatrix}
0 &\! G_{12} \!& \! \cdots \! &\!\! G_{1L} \!\\
\! G_{21} \!& 0 & \! \ddots \! &\!\!  \vdots \!\\
\vdots &\! \ddots \!& \! \ddots \! &\!\! G_{L-1 \ L} \!\\
\! G_{L1} \!&\! \cdots \!& \!\! G_{L \ L-1} \!\! &\!\! 0
\end{bmatrix} \!\!\!\!
\begin{bmatrix} \! w_1 \!\\[1pt]  \! w_2 \!\\[1pt] \! \vdots \!\\[1pt] \! w_L \! \end{bmatrix} \!\!\!
+ \!\! R \!\!
\begin{bmatrix} \! r_1 \!\\[1pt] \! r_2 \!\\[1pt] \! \vdots \!\\[1pt]  \! r_{K} \!\end{bmatrix}
\!\!\!+\!\!
H  \!\! \begin{bmatrix}\! e_1 \!\\[1pt] \! e_2 \!\\[1pt] \! \vdots \!\\[1pt] \! e_p\!\end{bmatrix} \!\!\!
\end{align*}
which results in the matrix equation:
\begin{align} \label{eq.dgsMatrix}
w = G w + R r + H e.
\end{align}
The network transfer function that maps the external signals $r$ and $e$ into the node signals $w$ is denoted by:
\begin{align}
T(q) &:= \begin{bmatrix} T_{wr}(q) & T_{we}(q) \end{bmatrix}, \ \mbox{with}& \\
T_{wr}(q) &:= \left( I-G(q)\right)^{-1}R(q), \ \mbox{and}& \label{eq:twr}\\
T_{we}(q) &:= \left( I-G(q)\right)^{-1}H(q).& \label{eq:twe}
\end{align}
As a shorthand notation we use
$
 	U(q) :=
 	\begin{bmatrix}
 		H(q) & R(q)
 	\end{bmatrix}.
$
The identification problem to be considered is the problem of identifying the network dynamics ($G, R, H, \Lambda$) on the basis of measured variables $w$ and $r$.

\begin{remark}
	The dynamic network formulation above is related to what has been called the {\it Dynamic Structure Function (DSF)} as considered for disturbance-free systems in \cite{adebayo2012dynamical,Yuan2011,Yuan2012}. 
\end{remark}

In order to arrive at a definition of network identifiability we need to specify a network model and  network model set.
\begin{definition}[network model]
\label{def1}
A network model of a network with $L$ nodes, and $K$ external excitation signals, with a noise process of rank $p \leq L$ is defined by the quadruple:
\[ M = (G,R,H,\Lambda) \]
with
\begin{itemize}
\item $G \in \R^{L \times L}(z)$, diagonal entries 0, all modules proper and stable\footnote{The assumption of having all modules stable is made in order to guarantee that $T_{we}$ (\ref{eq:twe}) is a stable spectral factor of the noise process that affects the node variables.};
\item $R \in \R^{L \times K}(z)$, proper;
\item $H \in \R^{L \times p}(z)$, stable, with a stable left  inverse, and 
$ \begin{bmatrix} 		I_p & 0 	\end{bmatrix} 	H(q)$ is monic.
\item    $\Lambda \in \R^{p\times p}$, $\Lambda > 0$;
\item the network is well-posed\footnote{This implies that all principal minors of $(I-G(\infty))^{-1}$ are nonzero.}  \cite{Dankers_diss}, with $(I-G)^{-1}$ proper and stable. \hfill $\Box$
\end{itemize}
\end{definition}
The noise model $H$ is defined to be non-square in the case of a rank-reduced noise ($p<L$).

\begin{definition}[network model set]
	\label{def2}
	A network model set for a network of $L$ nodes, $K$ external excitation signals, and a noise process of rank $p \leq L$, is defined as a set of parametrized matrix-valued functions:
	\[ \M := \left\{ M(\theta) = \bigl(G(q,\theta), R(q,\theta), H(q,\theta), \Lambda(\theta)\bigr), \theta \in \Theta \right\}, \]
	with all models $M(\theta)$ satisfying the properties as listed in Definition \ref{def1}.
	\hfill $\Box$
\end{definition}

A \emph{path} in the network is a sequence of modules.
More precisely there exists a path through nodes $w_{n_1}, \ldots, w_{n_k}$ if
\[ G_{n_1 n_2} G_{n_2 n_3} \cdots G_{n_{(k-1)} n_k} \neq 0. \]

\subsection{Identifiability}
Identification is usually performed on the basis of second-order properties of $w$ and $r$.
Therefore in \cite{weerts2018identifiability} network identifiability is defined as a property of the model set, on the basis of those second-order properties.
\begin{definition}[Network identifiability from \cite{weerts2018identifiability}]
\label{defif}
The network model set $\M$ is globally network identifiable at $M_0 :=M(\theta_0)$ if for all models $M(\theta_1) \in \M$,
\begin{equation} \label{equivTP}
		\left. \begin{array}{c} T_{wr}(q,\theta_1) = T_{wr}(q,\theta_0) \\ \Phi_{\bar v}(\omega,\theta_1) = \Phi_{\bar v}(\omega,\theta_0) \end{array} \right\}
		\Rightarrow
		M(\theta_1) = M(\theta_0),
\end{equation}
where $\Phi_{\bar v}$ is the spectrum of $\bar v(t) := w(t) - T_{wr}(q) r(t)$.
$\M$ is globally network identifiable if (\ref{equivTP}) holds for all $M_0 \in \M$.\hfill $\Box$
\end{definition}
Under some conditions on feedthrough in modules, the implication can be re-written.
\begin{proposition}[from \cite{weerts2018identifiability}]
	\label{prop:TTMM1}
	Let $\M$ be a network model that satisfies either
	\begin{itemize}
		\item all modules in $G(q,\theta)$ are strictly proper, or
		\item there are no algebraic loops\footnote{an algebraic loop is a path where $n_1=n_k$ and $\lim_{z\rightarrow\infty} G_{n_1 n_2}(z) G_{n_2 n_3}(z) \cdots G_{n_{(k-1)} n_k}(z) \neq 0$} and $\Lambda(\theta)$ is diagonal for all $\theta \in \Theta$.
	\end{itemize}
	Then $\M$ is globally network identifiable at $M_0 :=M(\theta_0)$ if for all models $M(\theta_1) \in \M$,
	\begin{eqnarray} \label{equivT}
		\lefteqn{\{T(q,\theta_1) = T(q,\theta_0)\}
		\Rightarrow} \\
		& & \ \ \ \{(G(\theta_1), R(\theta_1), H(\theta_1)) = (G(\theta_0),R(\theta_0), H(\theta_0))\}. \nonumber
	\end{eqnarray}
	Network model set $\M$ is globally network identifiable if (\ref{equivT}) holds for all $M_0 \in \M$.\hfill $\Box$
\end{proposition}
The results later in this paper also hold for for situations where modules are allowed to have algebraic loops; see \cite{weerts2018identifiability} for more details on the treatment of that situation.

Necessary and sufficient conditions for network identifiability can be formulated.
To this end we need to introduce some notation.
Considering row $j$, define $\check T_j$ as the transfer function from
signals $r$ and $e$ that are not input to parameterized transfers in $U_{j\star}(\theta)$,
to node signals $w$ that are input to parameterized transfers in $G_{j\star}(\theta)$.
The number of parameterized transfers in $G_{j\star}(\theta)$ and $U_{j\star}(\theta)$ are $\alpha_i$ and $\beta_i$ respectively. 
More formally, let $P_j$ be the permutation that gathers all parameterized modules on the left of $G_{j\star}(\theta)P_j$,
and let $Q_j$ be the permutation that gathers all non-parameterized transfers on the left of $U_{j\star}(\theta)Q_j$,
then
\begin{equation}
\label{eq:Ti}
	\check T_j(q,\theta) := \left[ I_{\alpha_j}\ \ 0 \right] P_j^{-1} T(q,\theta)Q_j\begin{bmatrix}I_{K+p-\beta_j} \\ 0  \end{bmatrix}.
\end{equation}

\begin{theorem}[Part of Theorem 2 from \cite{weerts2018identifiability}] \label{thm:theo3}
Let $\M$ satisfy the properties of Proposition \ref{prop:TTMM1}, and additionally satisfy:
\begin{itemize}
\item[a.] Every parametrized entry in the model $\{ M(z,\theta), \theta\in\Theta\}$ covers the set of all proper rational transfer functions;
\item[b.] All parametrized transfer functions in the model $M(z,\theta)$ are parametrized independently (i.e. there are no common parameters).
\end{itemize}
Then
	\begin{enumerate}
		\item $\M$ is globally network identifiable at $M(\theta_0)$ if and only if
		\begin{itemize}
			\item 	
			each row $j$ of the transfer function matrix $\begin{bmatrix} G(\theta) & U(\theta)	 \end{bmatrix}$ has at most $K+p$ parameterized entries, and
			\item
			for each $j$, $\check T_j(\theta_0)$ defined by (\ref{eq:Ti}) has full row rank.
		\end{itemize}
		\item $\M$ is globally network identifiable if and only if
		\begin{itemize}
			\item
			each row $j$ of the transfer function matrix $\begin{bmatrix} G(\theta) & U(\theta)	 \end{bmatrix}$ has at most $K+p$ parameterized entries, and
			\item
			for each $j$, $\check T_j(\theta)$ defined by (\ref{eq:Ti}) has full row rank for all $\theta \in \Theta$.	\hfill $\Box$		
		\end{itemize}
	\end{enumerate}	
\end{theorem}

\section{Extension to single-module identifiability}
In this section the identifiability of just part of the network, or a single module is treated.
To this end we formalize identifiability of particular properties of $M$ as suggested in \cite{weerts2018identifiability}.
First we define identifiability of a row of $M$, in order to evaluate identifiabilty around a certain node in a network, after which identifiability of a particular module is treated.
\begin{definition}
	\label{def:single}
	For network models that satisfy the conditions of Proposition \ref{prop:TTMM1},
	row $j$ of network model set $\M$ is globally network identifiable at $M_0 :=M(\theta_0)$ if for all models $M(\theta_1) \in \M$,
	\begin{equation} \label{equivTP2}
		\left . T(q,\theta_1) = T(q,\theta_0) \right \}
		\Rightarrow
		\left \{ \begin{array}{c} G_{j\star}(q,\theta_1) = G_{j\star}(q,\theta_0)\\ R_{j\star}(q,\theta_1) = R_{j\star}(q,\theta_0) \\ H_{j\star}(q,\theta_1) = H_{j\star}(q,\theta_0) \end{array}. \right .
	\end{equation}
	Row $j$ of network model set $\M$ is globally network identifiable if (\ref{equivTP2}) holds for all $M_0 \in \M$.\hfill $\Box$
\end{definition}
The conditions in Theorem \ref{thm:theo3} are formulated independently for each row, so it is straightforward to obtain conditions under which a specific row of $M$ is identifiable. 
\begin{corollary} \label{cor:node}
Let $\M$ be a network model set defined as in Theorem \ref{thm:theo3}, and $\check T_j (\theta)$ defined by (\ref{eq:Ti}),
	then
	\begin{enumerate}
		\item Row $j$ of network model set $\M$ is globally network identifiable at $M(\theta_0)$ if and only if
		\begin{itemize}
			\item 	
			row $j$ of the transfer function matrix $\begin{bmatrix} G(\theta) & U(\theta)	 \end{bmatrix}$ has at most $K+p$ parameterized entries, and
			\item
			$\check T_j(\theta_0)$  has full row rank.
		\end{itemize}
		\item Row $j$ of network model set $\M$ is globally network identifiable if and only if
		\begin{itemize}
			\item
			row $j$ of the transfer function matrix $\begin{bmatrix} G(\theta) & U(\theta)	 \end{bmatrix}$ has at most $K+p$ parameterized entries, and
			\item
			$\check T_j (\theta)$ has full row rank for all $\theta  \in  \Theta$.	\hfill $\Box$		
		\end{itemize}
	\end{enumerate}	
\end{corollary}

When we are interested in one specific module, then the above definition is still conservative.
It is possible that a module is identifiable, even when other modules of that row are not, 
which is illustrated by the following example.
\begin{example} \label{ex:singlemod}
	\begin{figure}[b]
		\centering
		\includegraphics[width=0.6\columnwidth,trim={1.2cm 0.55cm 0.6cm 0.6cm},clip]{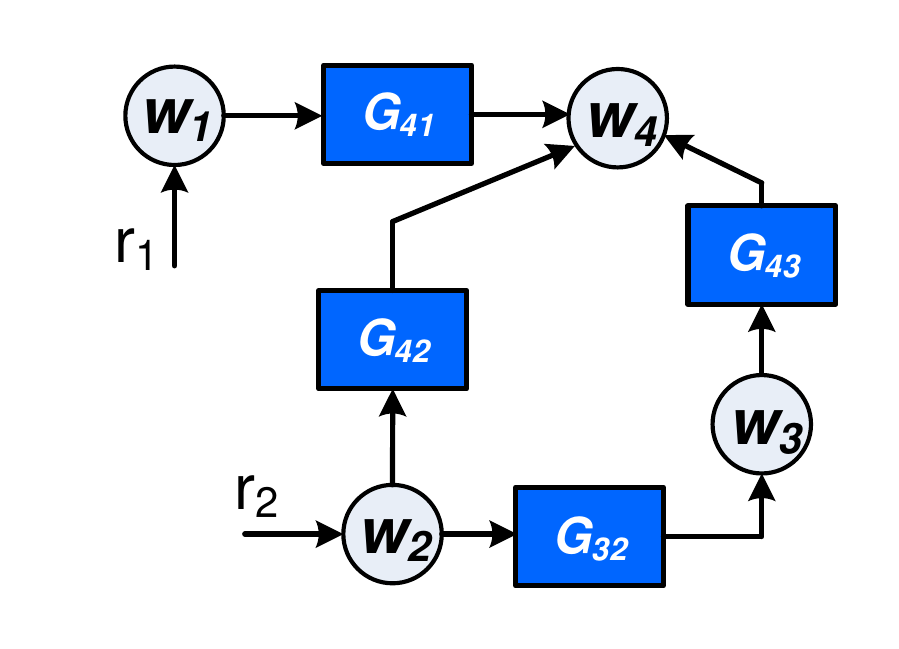}
		\caption{Example network model where some modules are identifiable.}
		\label{fig:1}
	\end{figure}
	Consider a set of network models of the topology shown in Figure \ref{fig:1}, described by
	\begin{equation}
		\begin{bmatrix} w_1\\ w_2\\ w_3 \\ w_4	\end{bmatrix}
		\! = \! 
		\begin{bmatrix} 0 & 0 & 0 & 0 \\ 0&0&0&0\\ 0 & G_{32} &0&0\\ G_{41} & G_{42} & G_{43} &0	\end{bmatrix}
		 \! 
		\begin{bmatrix} w_1\\ w_2\\ w_3 \\ w_4	\end{bmatrix}
		 \! + \! 
		\begin{bmatrix} 1&0\\0&1\\0&0\\0&0	\end{bmatrix}
		 \! 
		\begin{bmatrix} r_1 \\ r_2	\end{bmatrix}  \! ,
	\end{equation}
	where all modules $G_{ji}$ are parameterized.
	The response of the node variables is given by $w=Tr$ with
	\begin{equation}
		T = (I-G)^{-1}R= \begin{bmatrix}
			1&0\\0&1\\
			0&G_{32} \\ G_{41} & G_{42} + G_{32} G_{43}
		\end{bmatrix}.
	\end{equation}
	From $T$ we can directly determine $G_{41}$ and $G_{32}$,
	but we can not distinguish between $G_{42}$ and $G_{43}$.
	For node 4 there are three parameterized transfer functions, but only two excitations, so the model set is not globally network identifiable,
	but we can see that particular modules are identifiable.
	\hfill $\Box$
\end{example}

To define identifiability of a specific module  the implication \eqref{equivTP2} is made even more specific.
\begin{definition}
	\label{defif3}
	For network models that satisfy the conditions of Proposition \ref{prop:TTMM1},
	module $G_{ji}$ of network model set $\M$ is globally network identifiable at $M_0 :=M(\theta_0)$ if for all models $M(\theta_1) \in \M$,
	\begin{equation} \label{equivTP3}
		\{ T(q,\theta_1) = T(q,\theta_0) \}
		\Rightarrow
		\{ G_{ji}(q,\theta_1) = G_{ji}(q,\theta_0) \}.
	\end{equation}
	Module $G_{ji}$ of network model set $\M$ is globally network identifiable if (\ref{equivTP3}) holds for all $M_0 \in \M$. \hfill $\Box$
\end{definition}
It is obvious that identifiability of every module holds for every model set that is globally network identifiable.
However now the interesting question is whether the conditions can be relaxed, such that identifiability of a module is guaranteed, even when other modules are not identifiable.

In order to find identifiability conditions for a single module $G_{ji}$, assume without loss of generality that this module corresponds to the top row of $\check T_j$.
Then define $\check T_{j,(i,\star)}$ as the top row of $\check T_j$, and $\check T_{j,(-i,\star)}$ by
\begin{equation} \label{eq:chT}
	\check T_j(q,\theta) = \begin{bmatrix}
		\check T_{j,(i,\star)}(q,\theta) \\ 	\check T_{j,(-i,\star)}(q,\theta)
	\end{bmatrix}
\end{equation}
So $\check T_{j,(-i,\star)}$ is $\check T_j$ with the row corresponding to node $w_i$ removed.
The following Theorem now specifies necessary and sufficient conditions for the identifiability condition (\ref{equivTP3}).
\begin{theorem} \label{thm:theo2}
	Let $\M$ be a network model set defined as in Theorem \ref{thm:theo3},
	then
	\begin{enumerate}
		\item Module $G_{ji}$ of network model set $\M$ is globally network identifiable at $M(\theta_0)$ if and only if
		\begin{equation} \label{eq:singmod_cond0}
			\mathrm{rank}(\check T_j(\theta_0)) >  \mathrm{rank}(\check T_{j,(-i,\star)}(\theta_0)).
		\end{equation}		
		\item Module $G_{ji}$ of network model set $\M$ is globally network identifiable if and only if
		\begin{equation} \label{eq:singmod_cond}
			\mathrm{rank}(\check T_j(\theta)) >  \mathrm{rank}(\check T_{j,(-i,\star)}(\theta))
		\end{equation}	
		for all $\theta \in \Theta$.
		\hfill $\Box$	
	\end{enumerate}	
\end{theorem}
The proof is collected in the appendix.

The essential part of the theorem is that if the row of $\check T_j$ corresponding to node $w_i$ is a linear independent row, then the module is identifiable.
Note that there is no explicit requirement on the number of parameterized elements in Theorem \ref{thm:theo2}. 
We do not require uniqueness of all modules, so we can have fewer equations than unknowns.

\begin{example}[Example \ref{ex:singlemod} continued] \label{ex:cont}
	For node 4 there are three parameterized transfer functions, while there are only two excitations.
	The matrix to be evaluated is
	\begin{equation}
		\check T_4 =
		\begin{bmatrix}
			1 & 0 \\ 0& 1 \\ 0 & G_{32}
		\end{bmatrix}
	\end{equation}
	of dimension $3 \times 2$, so that it can never be full row rank.

	When evaluating the three modules on row 4 individually, only one row is linearly independent of the others.
	For module $G_{41}$ the matrices to be checked are
	\begin{equation}
		\check T_{4(1,\star)} = \begin{bmatrix}
		1&0
		\end{bmatrix},
		\quad
		\check T_{4(-1,\star)} = \begin{bmatrix}
			0&1\\0&G_{32}
		\end{bmatrix}.
	\end{equation}
	We can then clearly see  that $\check T_{4(1,\star)}$ is linearly independent of $\check T_{4(-1,\star)}$ and also that  \[\mathrm{rank}(\check T_{4(-1,\star)})=1 < \mathrm{rank}(\check T_{4})=2,\] such that the condition of  Theorem \ref{thm:theo2} is satisfied for $G_{41}$.
	
	It can be shown that rows 2 and 3 of $\check T_4$ are linearly dependent, and so $G_{42}$ and $G_{43}$ are both not identifiable.
\end{example}

\section{Path-based identifiability conditions}
In this section the rank conditions used for network identifiability are formulated as topology based conditions.
The core idea is that the rank of $T$ depends on the topology of the network.
We base our reasoning on concepts presented in \cite{bazanella2017identifiability}, 
where network identifiability is considered for situations where not all nodes are measured.
We adapt the identifiability definition in \cite{bazanella2017identifiability} to our problem setting, 
and then formulate topological conditions on the basis of \emph{disjoint paths}, which will be defined later.

The identifiability concept treated in \cite{bazanella2017identifiability} differs from Definition \ref{defif}, and we formulate the following definition in order to use their approach.
\begin{definition}[Generic network identifiability]
	\label{defif2}
	\hspace{1cm}
	\begin{itemize}
	\item	$\M$ is generically globally network identifiable if (\ref{equivT}) holds for \emph{almost} all $M_0 \in \M$.
	\item 	Row $j$ of network model set $\M$ is generically globally network identifiable if (\ref{equivTP2}) holds for almost all $M_0 \in \M$.
	\item 	Module $G_{ji}$ of network model set $\M$ is generically globally network identifiable if (\ref{equivTP3}) holds for almost all $M_0 \in \M$.
	\hfill $\Box$
	\end{itemize}
\end{definition}
The only difference between Definitions \ref{defif} and \ref{defif2} is the exception of a set of zero measure.
Implications that come with this different definitions are discussed in Section \ref{sec:discussion}.

The rank conditions of Theorems \ref{thm:theo3} and \ref{thm:theo2}, Corollary \ref{cor:node} can directly be formulated for the generic network identifiability.
\begin{corollary} \label{cor1}
The model set $\M$, row $j$ of model set $\M$, or module $G_{ji}$ of model set $\M$ is generically globally network identifiable by the conditions $2)$ of Theorem \ref{thm:theo3}, Corollary \ref{cor:node}, Theorem \ref{thm:theo2} respectively upon replacing the phrase ``for all $\theta \in \Theta$" by ``for \emph{almost} all $\theta \in \Theta$".
	\hfill $\Box$
\end{corollary}

The proof is a trivial extension of the proof of Theorem \ref{thm:theo3} found in \cite{weerts2018identifiability} and the proof of Theorem \ref{thm:theo2}.

In order to formulate topological conditions under which a model set is generically globally network identifiable, the notion of disjoint paths is introduced following the approach in \cite{bazanella2017identifiability} and the definition in \cite{Woude1991}.
Consider two paths in the network, then we can say that these two paths are disjoint if they have no common nodes, including their start and end nodes or excitations.
For a set of $l$ paths, these paths are disjoint if every pair of paths is disjoint.

Essentially what this means is that if there exists a set of disjoint paths from some excitations $r_k$, $e_l$ to some nodes $w_i$, then every one of those nodes has 'its own' source of excitation.
Note that when two paths are disjoint, there may still exist modules that connect the nodes in the paths, and there may exist loops around the nodes.

In \cite{Woude1991} the notion of a set of disjoint paths is connected to the rank of a transfer matrix.
This is defined on the basis of state-space systems in the following way.
A parameterized state-space system is defined with matrices $A,B,C$, and the open-loop transfer from input to output is defined as $C(sI-A)^{-1}B$.
Then the \emph{generic rank} of the transfer matrix $C(sI-A)^{-1}B$ is defined as the rank that $C(sI-A)^{-1}B$ has for almost all parameters.
From the paper then the relation between rank and disjoint paths is formulated.
\begin{theorem}[Theorem 2 from \cite{Woude1991}] \label{thm:generic}
	Let $G_\Sigma$ be the graph corresponding to the state-space system
	\begin{equation}
		\begin{split}
			\dot x &= Ax + Bu \\
			y &= Cx.
		\end{split}
	\end{equation}
	The maximum number of  disjoint paths in $G_\Sigma$ from signals in $u$ to signals in $y$ is equal to the generic rank of $C(sI-A)^{-1}B$.\hfill $\Box$	
\end{theorem}

This state-space representation is very similar to the network representation.
If we take a network with $C=I$ and first order modules, then this network is equivalent to the state-space model.
In that case we have that
\begin{equation}
 	C(sI-A)^{-1}B = (I-G)^{-1} U = T,
\end{equation}
where the graph $G_\Sigma$ has the same topology as $G$.
When the order of the modules of $G$ is allowed to be greater than 1, then this has no effect on the topology of $G$, and also the generic rank of $T$ does not depend on the order of the modules in the network.

\begin{proposition} \label{prop:path}
	Let $\M$ be a set of network models $M$ with strictly proper modules.
	Let $\U$ be a set of excitations, i.e. a set of some $r_k$ and $e_l$, and let $\Y$ be a set of nodes $w_i$.
	The maximum number of disjoint paths in $M$ from excitations in $\U$ to nodes in $\Y$ is equal to the generic rank of
	the transfer $T_{\Y\U}(q,\theta)$ from excitations in $\U$ to nodes  in $\Y$.
	\hfill $\Box$	
\end{proposition}

Using Proposition \ref{prop:path} the conditions on $\check T_j$ of Corollary \ref{cor1} can be explained using disjoint sets.
The $\check T_j$ is the transfers from external signals $r_k$, $e_l$ that are input to non-parameterized transfers, to node signals $w_i$ that are input to parametrized modules that map to $w_j$.
So then we know that the generic row rank of $\check T_j$ can be checked by checking whether there are a sufficient number of disjoint paths from excitations in $\U_j$ to nodes in $\Y_j$.
\begin{proposition} \label{prop:path_cond}
	Let $\Y_j$ be the set of nodes $w_k$ which are an input to a $G_{jk}$ that is parameterized.
	Let $\alpha_j$ be the number of parameterized modules that map into node $j$, i.e. the cardinality of $\Y_j$.
	Let $\U_j$ be the set of excitations $r_k$, $e_l$ that are \emph{not} an input to a $R_{jk}$, $H_{jl}$ that is parameterized.
	The three conditions on matrix rank referred to in Corollary \ref{cor1} are equivalently formulated as:
	\begin{enumerate}
		\item For each $j$, there is a set of $\alpha_j$ disjoint paths from excitations in $\U_j$ to nodes in $\Y_j$;
		\item There is a set of $\alpha_j$ disjoint paths from excitations in $\U_j$ to nodes in $\Y_j$;
		\item For the module of interest $G_{ji}$, let $\bar \Y_j = \Y_j \setminus w_i$.
		There is a set $\mathcal P$ of the maximum number of disjoint paths from excitations in $\U_j$ to nodes in $\bar \Y_j$, and an additional path from excitations in $\U_j$ to $w_i$ that is disjoint to the paths in $\mathcal P$.
			\hfill $\Box$
	\end{enumerate}
\end{proposition}
In order to satisfy condition 1) or 2) there is an implicit requirement on the number of available excitations, which is directly related to the maximum number of parameterized elements in conditions 1) and 2) of Corollary \ref{cor1}.
For condition 3) there is no minimum number of excitations, but there is the implicit requirement that there is a 'surplus' excitation that can form a disjoint path to the module of interest.

In order to check the conditions of Proposition \ref{prop:path_cond}, all that must be done is  check which transfer functions are parameterized, and check whether the necessary paths are present in the network.
This is illustrated in an example.
\begin{example}[Example \ref{ex:cont} continued]
	Now using the topology based condition the identifiability of modules is checked.
	
	In order to check the identifiability of modules that map into node $w_4$ we see that $\Y_4=\{w_1,w_2,w_3\}$, so $\alpha_4=3$.
	There are only two excitations present in the network which are not an input to $w_4$, $\U_4=\{r_1,r_2\}$, so immediately we know that there can not be 3 disjoint paths from excitation to $\Y_4$, and that row 4 of $M$ is not generically network identifiable.

	For identifiability of single modules we see that there are two disjoint paths from $\U_4$ to $\Y_4$.
	Then for module $G_{42}$ we see there are two disjoint paths from $\U_4$ to $\bar \Y_4=\{w_1,w_3\}$, so $G_{42}$ is not generically network identifiable.
	However  for module $G_{41}$ there is just one disjoint path from $\U_4$ to $\bar \Y_4=\{w_2,w_3\}$, so $G_{41}$ is generically network identifiable.
	Basically when $w_1$ is removed, there is a surplus excitation $r_1$ that can not form a disjoint path to $w_3$.
\end{example}

\section{Discussion on definition of identifiability} \label{sec:discussion}
Path-based conditions are based on generic rank, and not 'normal' rank.
The difference between the two definitions of identifiability is the exception of a zero-measure set of models, so network identifiability is more strict than generic network identifiability.
When one model in $\M$ is not identifiable, then $\M$ is not network identifiable, but it can be generically network identifiable.
In order to understand the difference between the definitions, we need to understand which models cause the difference, and whether those models are important.
An illustration of this is given in an example.

\begin{example} \label{ex:loop}
	\begin{figure}[htb]
		\centering
		\includegraphics[width=0.6\columnwidth,trim={0.7cm 0.6cm 0.7cm 0.8cm},clip]{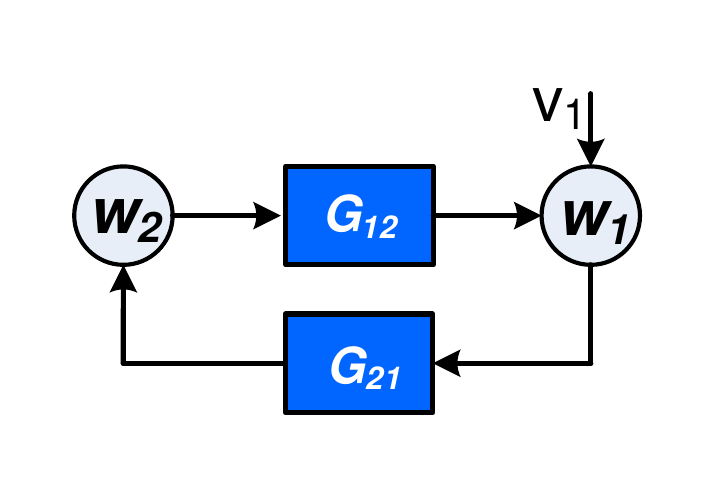}
		\caption{A closed-loop network representing a set of models.}
		\label{fig:cl}
	\end{figure}
	Suppose we have a parameterized set of models as depicted in Figure \ref{fig:cl}, with
	\begin{equation*}
 		G = \begin{bmatrix}
 			0 & G_{12} \\ G_{21} & 0
 		\end{bmatrix},
 		\quad
 		H = \begin{bmatrix}
 			1 \\ 0
 		\end{bmatrix},
 		\quad
 		T^0 = \begin{bmatrix}
 			\frac{1}{1-G_{12}^0 G_{21}^0} \vspace{3pt} \\ \frac{G_{21}^0}{1-G^0_{12}G^0_{21}}
 		\end{bmatrix}.
	\end{equation*}
	Now the identifiability question is whether $G_{12}$ and $G_{21}$ can be uniquely determined from
	\begin{equation} \label{eq:ex1}
	 	 \begin{bmatrix}
 			0 & G_{12} \\ G_{21} & 0
 		\end{bmatrix}
 		\begin{bmatrix}
 			\frac{1}{1-G_{12}^0 G_{21}^0} \\ \frac{G_{21}^0}{1-G^0_{12}G^0_{21}}
 		\end{bmatrix}
 		=
 		\begin{bmatrix}
 			1 \\ 0
 		\end{bmatrix}.
	\end{equation}
	When $G_{21}^0 \ne 0$, then $T^0$ has two non-zero entries, and \eqref{eq:ex1} consists of two independent equations with two unknowns.
	However if $G_{21}^0 = 0$, then $T^0$ has a 0 entry, and \eqref{eq:ex1} has one trivial equation, such that $G_{12}$ can not be determined uniquely.
	If we have a-priori knowledge that $G_{21}^0 \ne 0$, then $G_{12}$ can be determined uniquely, and	we want to classify the model set as identifiable.
	However in a topology detection situation we would like to determine whether $G_{12}$ and $G_{21}$ are zero or non-zero, so the possibility that $G_{21}^0 = 0$ must be taken into account.
	Precisely in that situation the $G_{12}$ can not be determined from data, and we want to classify the model set as non-identifiable.
 	\hfill $\Box$
\end{example}

Typically there are three possible objectives in a network identification problem: topology detection, identification of all modules, or identification of a single module.
For problems where modules may be 0, such as in topology detection problems, we have to be able to distinguish between all possible models, even when modules are 0.
So for those problems global network identifiability is the desired concept.

\section{Conclusions}
The notion of network identifiability has been extended to cover the case of single-module identifiability.
Necessary and sufficient conditions for single module identifiability have been obtained, and it has been shown that when considering a generic version of the identifiability concept, the necessary and sufficient conditions can be reformulated in terms of path-based conditions that can simply be verified on the basis of the network topology.

\appendix
\subsection{Proof of Theorem \ref{thm:theo2}}
The left hand side of the implication (\ref{equivTP3}) can be written as
	\begin{equation} \label{eq:row}
		(I-G(\theta)) T = U(\theta),
	\end{equation}
	where we use shorthand notation $T = T(\theta_0)$, $G(\theta) = G(\theta_1)$ and $U(\theta)=U(\theta_1)$.
By inserting the permutation matrices $P$ and $Q$ as in \eqref{eq:Ti} we obtain for row $j$:
\begin{equation}
(I-G(\theta))_{j\star} PP^{-1}TQ = U_{j \star}(\theta)Q
\end{equation}
leading to
	\begin{equation}\label{eq11}
		(I-G(\theta))_{j\star}^{(1)} T^{(1)}_j + (I-G)_{j \star}^{(2)} T^{(2)}_j = \begin{bmatrix} U_{j \star}^{(1)} &  U(\theta)_{j\star}^{(2)} \end{bmatrix},
	\end{equation}
with $P^{-1}TQ = \begin{bmatrix} T^{(1)}_j \\ T^{(2)}_j \end{bmatrix}$.
Note that $\check T_j =  T_j^{(1)}\begin{bmatrix} I_{K+p-\beta} \\ 0 \end{bmatrix}$.
The right-hand block in (\ref{eq11}) corresponding to $U(\theta)_{j\star}^{(2)}$ does not add to the uniqueness of the module of interest since it is fully parameterized (conditions \emph{a,b} of Theorem \ref{thm:theo3}), so equivalently we can consider
\begin{equation}\label{eq:perm}
	(I-G(\theta))_{j\star}^{(1)}\breve T_j + \rho = U_{j \star}^{(1)},
\end{equation}	
with $\rho$ the left $1\times (K+p-\beta)$ block of $(I-G)_{j\star}^{(2)}T_j^{(2)}$.
Now since $\rho$ and $U_{j \star}^{(1)}$ are independent of $\theta$ we have that $(I-G(\theta))_{ji}^{(1)}$  is uniquely specified if and only if $(I-G(\theta))_{ji}^{(1)}$ is uniquely specified in the left-nullspace of $\check T_j$.

\textbf{Sufficiency:}
	\\
Define some transfer matrix $X(q)$ of dimension $(K+p-\beta) \times 1$ with the following properties:
\begin{itemize}
	\item $\check T_{j(-i,\star)}(q,\theta_0) X(q) = 0$, and
	\item $\check T_{j(i,\star)}(q,\theta_0) X(q) \neq 0$,
\end{itemize}
where $\check T_{j(-i,\star)}$ and $\check T_{j(i,\star)}$ are defined in \eqref{eq:chT}.
This $X$ exists because condition \eqref{eq:singmod_cond0} requires that $\check T_{j(-i,\star)}(q,\theta_0)$ is not full column rank, and condition \eqref{eq:singmod_cond0} implies that $\check T_{j(i,\star)}(q,\theta_0)$ is linearly independent from the rows of $\check T_{j(-i,\star)}(q,\theta_0)$.
Now define an  $(K+p-\beta) \times (K+p-\beta)$ full rank transfer matrix $Z(q)$ which has $X$ as its first column.
Then \eqref{eq:perm} can be post-multiplied with $Z$ to obtain an equivalent set of equations,
leaving the set of solutions for $G_{ji}$ invariant.
The first column of $\check T_j Z$ is
\begin{equation} 
	\check T_j(q,\theta_0) X(q) = \begin{bmatrix} \check T_{j(i,\star)}(q,\theta_0) X(q) \\ 0 \end{bmatrix}, 
\end{equation}
such that, for this choice of $Z$, $G_{ji}$ can be uniquely determined from
\begin{equation}
 	(I-G(\theta))_{j\star}^{(1)} \begin{bmatrix} \check T_{j(i,\star)} X \\ 0 \end{bmatrix} 
 	= 
 	(U_{j\star}^{(1)} - \rho) X.
\end{equation}
If $G_{ji}$ is unique for this particular choice of $Z$, it must be unique in the original problem also. 

\textbf{Necessity:}
\\
The converse of condition \eqref{eq:singmod_cond0} is that rank$\big( \check T_{j}(q,\theta_0) \big ) = $ rank$\big( \check T_{j(-i,\star)}(q,\theta_0) \big )$.
In this case the row of $\check T_{j}(q,\theta_0)$ corresponding to $G_{ji}(\theta)$ is linearly dependent on other rows of $\check T_{j}(q,\theta_0)$.
When $\check T_{j(i,\star)}$ is linearly dependent on another row $\check T_{j(k,\star)}$, an equation equivalent to \eqref{eq:perm} can be created where the element $G_{j1}$ and row $\check T_{j(i,\star)}$ are deleted, and where $( G_{ji} F + G_{jk}  )$ replaces $G_{jk}$,
such that $G_{ji}$ can not uniquely be distinguished.

\textbf{Proof of situation (2): For all $\theta \in \Theta$:}\\
	For every $\theta \in \Theta$ we can construct $T(\theta)$ with related $\check T_j(\theta)$.
	If condition \eqref{eq:singmod_cond0} applies for every model as stated by condition \eqref{eq:singmod_cond},
	then the reasoning as presented before fully applies to every model.
	If for some $\theta \in \Theta$ the condition \eqref{eq:singmod_cond0} is not met, there exists a model in the model set which is not identifiable, and hence the model set is not globally network identifiable in $\M$.
\hfill $\Box$


\bibliography{Library}{}

\begin{thebibliography}{10}
\providecommand{\url}[1]{#1}
\csname url@samestyle\endcsname
\providecommand{\newblock}{\relax}
\providecommand{\bibinfo}[2]{#2}
\providecommand{\BIBentrySTDinterwordspacing}{\spaceskip=0pt\relax}
\providecommand{\BIBentryALTinterwordstretchfactor}{4}
\providecommand{\BIBentryALTinterwordspacing}{\spaceskip=\fontdimen2\font plus
\BIBentryALTinterwordstretchfactor\fontdimen3\font minus
  \fontdimen4\font\relax}
\providecommand{\BIBforeignlanguage}[2]{{%
\expandafter\ifx\csname l@#1\endcsname\relax
\typeout{** WARNING: IEEEtran.bst: No hyphenation pattern has been}%
\typeout{** loaded for the language `#1'. Using the pattern for}%
\typeout{** the default language instead.}%
\else
\language=\csname l@#1\endcsname
\fi
#2}}
\providecommand{\BIBdecl}{\relax}
\BIBdecl

\bibitem{Goncalves08}
J.~Gon\c{c}alves and S.~Warnick, ``Necessary and sufficient conditions for
  dynamical structure reconstruction of {LTI} networks,'' \emph{IEEE Trans.
  Automatic Control}, vol.~53, no.~7, pp. 1670--1674, Aug. 2008.

\bibitem{Materassi2010}
D.~Materassi and G.~Innocenti, ``Topological identification in networks of
  dynamical systems,'' \emph{IEEE Trans. on Automatic Control}, vol.~55, no.~8,
  pp. 1860--1871, 2010.

\bibitem{Yuan2011}
Y.~Yuan, G.~B. Stan, S.~Warnick, and J.~Gon\c{c}alves, ``Robust dynamical
  network structure reconstruction,'' \emph{Automatica}, vol.~47, no.~6, pp.
  1230--1235, 2011.

\bibitem{ChiusoPAuto2012}
A.~Chiuso and G.~Pillonetto, ``A {B}ayesian approach to sparse dynamic network
  identification,'' \emph{Automatica}, vol.~48, no.~8, pp. 1553–--1565, 2012.

\bibitem{VandenHof&Dankers&etal:13}
P.~M.~J. Van~den Hof, A.~G. Dankers, P.~S.~C. Heuberger, and X.~Bombois,
  ``Identification of dynamic models in complex networks with prediction error
  methods - basic methods for consistent module estimates,'' \emph{Automatica},
  vol.~49, no.~10, pp. 2994--3006, 2013.

\bibitem{Dankers_diss}
A.~G. Dankers, ``System identification in dynamic networks,'' Ph.D.
  dissertation, Delft University of Technology, 2014.

\bibitem{weerts2018identifiability}
H.~H.~M. Weerts, P.~M.~J. Van~den Hof, and A.~G. Dankers, ``Identifiability of
  linear dynamic networks,'' \emph{Automatica}, vol.~89, pp. 247--258, 2018.

\bibitem{bazanella2017identifiability}
A.~Bazanella, M.~Gevers, J.~Hendrickx, and A.~Parraga, ``Identifiability of
  dynamical networks: which nodes need be measured?'' in \emph{Proc. 56th IEEE
  Conf. Decision and Control (CDC 2017)}, 2017, pp. 5870--5875.

\bibitem{adebayo2012dynamical}
J.~Adebayo, T.~Southwick, V.~Chetty, E.~Yeung, Y.~Yuan, J.~Gon\c{c}alves,
  J.~Grose, J.~Prince, G.-B. Stan, and S.~Warnick, ``Dynamical structure
  function identifiability conditions enabling signal structure
  reconstruction,'' in \emph{Decision and Control (CDC), 2012 IEEE 51st Annual
  Conf. on}.\hskip 1em plus 0.5em minus 0.4em\relax IEEE, 2012, pp. 4635--4641.

\bibitem{Yuan2012}
Y.~Yuan, ``Decentralised network prediction and reconstruction algorithms,''
  {PhD} dissertation, University of Cambridge, 2012.

\bibitem{Soderstrom&Stoica:89}
T.~S\"{o}derstr\"{o}m and P.~Stoica, \emph{System Identification}.\hskip 1em
  plus 0.5em minus 0.4em\relax Hemel Hempstead, UK: Prentice Hall, 1989.

\bibitem{Woude1991}
J.~van~der Woude, ``A graph-theoreric characterization for the rank of the
  transfer matrix of a structured system,'' \emph{Mathematics of Control,
  Signals, and Systems}, vol.~4, pp. 33--40, 1991.

\end{thebibliography}
\bibliographystyle{IEEEtran}

\end{document}